\documentclass[12pt]{article}

\usepackage{sbc-template}
\usepackage{appendix}
\usepackage[section]{placeins}
\usepackage{graphicx,url}
\usepackage[caption=false]{subfig}
\usepackage{multirow}
\usepackage[portuguese,linesnumbered,ruled,vlined]{algorithm2e}
\usepackage{diagbox}
\usepackage[brazil]{babel}
\usepackage{color}
\usepackage{amsmath,amssymb,amsfonts,amsthm}
\graphicspath{{Histogramas/}{./}}
\newcommand\nocell[1]{\multicolumn{#1}{c|}{}}
\usepackage{bm} % for bold greek letters
\newcommand*{\bx}{\mathbf{x}}
\newcommand*{\bX}{\mathbf{X}}

\newcommand*{\bbeta}{\bm{\beta}}
\newcommand*{\bZ}{\mathbf{Z}}
\newcommand*{\bT}{\mathbf{T}}
\newcommand*{\bzero}{\mathbf{0}}
\newcommand*{\bone}{\mathbf{1}}
\newcommand*{\MSE}{\textrm{MSE}}
\newcommand*{\var}{\textrm{var}}

\newcommand{\techreport}[2]{#1}

\newtheoremstyle{exampstyle}
  {5pt} % Space above
  {5pt} % Space below
  {\it} % Body font
  {} % Indent amount
  {\bfseries} % Theorem head font
  {.} % Punctuation after theorem head
  {.5em} % Space after theorem head
  {} % Theorem head spec (can be left empty, meaning `normal')
\theoremstyle{exampstyle} \newtheorem{theorem}{Teorema}

\usepackage[utf8]{inputenc}

\newcommand{\MODEL}{$\tau$-Exposure}

\newcommand{\p}{\phantom{*}}

\title{Modelos de Resposta para Experimentos Randomizados\\em Redes Sociais de Larga Escala\footnote{O presente trabalho foi realizado com apoio do CNPq, CAPES e FAPEMIG.}}

\author{Francisco Galuppo Azevedo, Bruno Demattos Nogueira,\\ Fabricio Murai, Ana Paula Couto da Silva}

\address{Departamento de Ciência da Computação\\
   Universidade Federal de Minas Gerais (UFMG) -- Belo Horizonte, MG -- Brazil
\email{\{franciscogaluppo, bruno.demattos, murai, ana.coutosilva\}@dcc.ufmg.br}
}
\begin{document}
\maketitle
\begin{abstract}
A/B tests are randomized experiments frequently used by companies that offer services on the Web for assessing the impact of new features. During an experiment, each user is randomly redirected to one of two versions of the website, called treatments. Several response models were proposed to describe the behavior of a user in a social network website, where the treatment assigned to her neighbors must be taken into account. However, there is no consensus as to which model should be applied to a given dataset. In this work, we propose a new response model, derive theoretical limits for the estimation error of several models, and obtain empirical results for cases where the response model was misspecified.
\end{abstract}

\begin{resumo}
Testes A/B são experimentos randomizados muito utilizados por empresas que oferecem serviços na Web para avaliar o impacto de novas funcionalidades. Durante um experimento, cada usuário é redirecionado aleatoriamente para uma de duas versões do site, chamadas tratamentos. Diversos modelos de resposta foram propostos para descrever o comportamento de um usuário em sites de redes sociais, onde o tratamento atribuído aos seus vizinhos deve ser considerado. Porém, não há consenso sobre qual modelo deve ser aplicado a um conjunto de dados. Neste trabalho, propomos um novo modelo de resposta, derivamos limites teóricos para o erro de estimação de diversos modelos, e obtemos resultados empíricos para o caso de especificação incorreta do modelo.
\end{resumo}

\section{Introdução}

Há alguns anos grandes empresas que oferecem serviços na Web (como Amazon, e-Bay, Facebook, Google e LinkedIn) vêm percebendo a importância de se conduzir pesquisas sobre a experiência dos usuários para a tomada de decisões, a nível de desenvolvimento (p.\ ex., mudanças no funil de compra ou no layout do site) e também a nível de negócio (p.\ ex., novas funcionalidades e diferenciais em relação aos concorrentes) \cite{kohavi2013online}.
%Para este fim, é necessária a criação de plataformas que permitam lançar, automaticamente, uma mudança ou funcionalidade no site e então coletar dados dos usuários a fim de testar se esta teve um impacto positivo.
Este impacto é quantificado através de medidas de interesse, tais como a fração de usuários que retorna ao site, o número médio de cliques por usuário, o lucro obtido através de anúncios no site etc. Para cada medida, deve-se estimar o efeito médio do tratamento, conhecido como {\it average treatment effect} (ATE).

Dentre as técnicas existentes, testes A/B se destacam como uma das mais proeminentes devido a sua capacidade de quantificar mudanças comportamentais objetivamente, calcular resultados e determinar a significância estatística de forma automática (sem que seja necessário feedback explícito do usuário) \cite{kohavi2009controlled}. Testes A/B consistem em experimentos randomizados com duas variantes de um tratamento. Usuários que visitam um website são aleatoriamente atribuídos ao grupo de controle (versão atual) ou ao grupo de tratamento (versão sendo testada).

Os testes A/B têm como premissa a chamada {\it Stable Unit Treatment Value Assumption} (SUTVA). A SUTVA é a suposição de que não há interferência entre os indivíduos do experimento, isto é, que o comportamento de um indivíduo depende apenas do tratamento que lhe foi atribuído. Enquanto esta suposição é válida para testes clínicos envolvendo placebos e medicamentos, experimentos em redes sociais normalmente violam esta condição (efeito conhecido como {\it spillover}) \cite{xu2015infrastructure}. Considere, por exemplo, um experimento para avaliar a criação de um chat para troca de mensagens entre usuários. Mesmo que as diferenças entre os grupos de controle e de tratamento não sejam significativas quando apenas 20\% dos usuários tem acesso à funcionalidade, não podemos afirmar que essa conclusão continuaria válida caso 100\% dos usuários pudessem usar o chat.
    
% ideal case. typical models for estimation. absence of ground-truth. test methods on synthetic data. benchmark: set of models for synthetic data. model mis-specification.
Idealmente, gostaríamos de comparar os resultados de dois experimentos simultâneos: no primeiro, toda a população está no grupo de controle e, no segundo, todos estão no grupo de tratamento. Desta forma, mesmo que a SUTVA fosse violada, conseguiríamos estimar o ATE. Como isso não é possível, para se estimar o ATE, é preciso assumir um modelo que descreve como a resposta de um indivíduo varia conforme o tratamento que é atribuído a ele e ao restante da população. No caso de redes sociais, assume-se que a resposta de um indivíduo é influenciada apenas pelos seus relacionamentos diretos.

Modelos de resposta são importantes não só para estimar o ATE, como também para comparar vários métodos de estimação quanto a sua acurácia. Como a função de resposta em um experimento real é desconhecida, a comparação entre métodos é feita a partir de dados sintéticos, gerados utilizando-se modelos de resposta. Tradicionalmente, os modelos de resposta utilizados são diferentes daqueles assumidos pelos métodos propostos. Contudo, para que os resultados em relação ao erro de estimação sejam colocados em perspectiva, é fundamental conhecer os limites teóricos do erro de estimação, caso o modelo correto fosse empregado.

Este trabalho possui três contribuições principais. A primeira consiste em propor um novo modelo de resposta, chamado $\tau$-exposure, baseado em limiar de exposição. Neste modelo um indivíduo é considerado ``exposto'' se ele recebe o controle (tratamento) e uma fração de seus vizinhos maior que $\tau$ também recebe controle (tratamento). A segunda contribuição é a derivação analítica dos limites inferiores para o erro de estimação quando o modelo correto de interferência é utilizado. Finalmente, iremos utilizar diversos modelos de resposta para realizar um estudo empírico do erro resultante ao se especificar o modelo incorreto durante a estimação.

\section{Trabalhos relacionados}

O estudo da resposta a um tratamento na presença de interações sociais foi formalizado em~\cite{Manski:2013fb}. O autor considera a existência de grupos de indivíduos conectados, cujas respostas são fortemente dependentes. Na classe mais geral de modelos, conhecida como {\it constant treatment response}, o espaço de todos os mapeamentos possíveis em grupos de controle e tratamento é particionado em ``tratamentos efetivos'' para cada indivíduo. Dois mapeamentos quaisquer dentro da mesma partição dão origem a mesma função de resposta para um indivíduo.

Com o auxílio de um grafo é possível definir modelos de resposta computacionalmente tratáveis. Estes modelos fazem parte da classe {\it neighborhood treatment response}. Por exemplo, em~\cite{Backstrom:2011hs} os autores assumem que um indivíduo é considerado ``exposto'' ao tratamento se ele e pelo menos $k$ de seus vizinhos recebem o tratamento. Além desse modelo, os autores de~\cite{Ugander:2013vc} consideram modelos onde um indivíduo é considerado exposto se todos os (ou uma fração maior que $q$ dos) seus vizinhos estão sujeitos ao mesmo tratamento que ele. Apesar de apresentarem resultados empíricos, nenhum destes trabalhos estuda os limites teóricos de estimação.

Em~\cite{Gui:2015ed}, os autores propõem um modelo linear para a resposta de um nodo em função do tratamento dele e dos seus vizinhos. Para reduzir o impacto de supor incorretamente tal relação linear, os autores usam randomização a nível de cluster. Por outro lado, o modelo, que produz respostas reais, é aplicado a respostas binárias sem que haja um estudo sobre o impacto desta inconsistência. Embora diversos modelos de resposta tenham sido propostos para capturar a interferência, não existe uma regra para escolher aquele que melhor representa o conjunto de dados, tampouco um entendimento do erro que a especificação incorreta pode ocasionar. Atualmente, existe apenas um método que permite testar se a SUTVA é válida \cite{Saveski2017}, assim como calcular a probabilidade de erro do Tipo I (rejeitar a SUTVA quando ela é válida).   

Além dos modelos de resposta onde há interferência entre indivíduos conectados, existem também modelos de homofilia, onde a resposta dos nodos conectados tende a ser similar quando sujeitos ao mesmo tratamento~\cite{Basse:2015tc}. Contudo, o foco do presente trabalho é em modelos de resposta onde há {\it spillover}.

%Em \cite{Moussaid2018} propõe-se um modelo para discussões em pequenos grupos, em que todos se relacionam e se influenciam, a fim de estudar o erro entre entre a estimativa para a qual a discussão converge e o valor verdadeiro. Porém, não é óbvio como generalizar os resultados para redes com mais de três participantes ou para aplicações mais gerais.

\section{Modelos de resposta}\label{sec:response}

Descrevemos a seguir a notação usada nos modelos de resposta. Seja cada unidade (ou indíviduo) do experimento indexada por $i \in 1, \ldots, N$. Defina $\mathbf{Z} \in \mathcal{Z}^N$ e $\mathbf{Y} \in \mathcal{Y}^N$ como sendo os vetores de tratamento e de resposta, onde $Z_i \in \mathcal{Z}$ e $Y_i \in \mathcal{Y}$ são, respectivamente, o tratamento atribuído ao e a resposta do usuário $i$.
%Define $f_i(\mathbf{Z})$ and $g_i(\mathbf{Y})$ as unit-specific functions computed over the treatment assignment $\mathbf{Z}$ and the response vector $\mathbf{Y}$, respectively. When units are connected in a network topology, typical examples of such functions include the fraction or number of neighbors that exhibit a given treatment or response.
Defina $g_i(\mathbf{Z})$ como sendo uma função específica à unidade $i$, calculada sobre a atribuição de tratamento $\mathbf{Z}$. Quando as unidades estão conectadas segundo uma topologia de rede, exemplo típicos de tais funções são a fração ou o número de vizinhos a que foram atribuídos um dado tratamento. Defina o vetor coluna $\bx_i = [1, Z_i, g_i(\mathbf{Z})]^\top$. Em geral, existe um componente estocástico $\epsilon_i$ associado à unidade $i$, amostrado independentemente para $i=1,\ldots,N$ a partir da distribuição Gaussiana com média zero e variância desconhecida $\sigma^2$. A seguir, iremos considerar apenas experimentos envolvendo dois tratamentos (testes A/B), logo $Z_i \in \mathcal{Z} \equiv \{0,1\}$.

\subsection{Modelo linear de resposta}

Quando as respostas são números reais (i.e., $Y_i \in \mathcal{Y} \equiv \mathbb{R}$), o modelo linear de resposta costuma ser utilizado:
\begin{eqnarray}
Y_i & = & \beta_0 + \beta_1 Z_i + \beta_2 g_i(\mathbf{Z}) + \epsilon_i \nonumber \\
 & = & \bx_i^\top \bbeta \ + \epsilon_i, \label{eq:linear}
\end{eqnarray}
onde $\bm{\beta} = (\beta_0,\beta_1,\beta_2)$ são parâmetros do modelo.

\subsection{Modelo probit}

Uma forma de se produzir respostas binárias (i.e., $Y_i \in \mathcal{Y} \equiv \{0,1\}$) a partir de uma função real (p. ex., $\bx_i^\top \bbeta \ + \epsilon_i$) consiste em usar o sinal da função para determinar a resposta. Este é caso do modelo probit:
\begin{eqnarray}
\Pr(Y_i = 1|\bx_i) & = &  \Pr(\bx_i^\top \bbeta + \epsilon_i > 0) \nonumber \\
& = & \Pr(\epsilon_i > -\bx_i^\top\bbeta ) \nonumber \\
& = & \Pr(\epsilon_i < \bx_i^\top\bbeta ), \nonumber \\
& = & \Phi(\bx_i^\top\bbeta) \label{eq:probit}
\end{eqnarray}
onde $\Phi(.)$ é a função de distribuição cumulativa da Normal de média $0$ e variância $\sigma^2$.

\subsection{Modelo logístico}

Outra forma de se produzir respostas binárias é mapear o valor de uma função real $f(.)$ em uma probabilidade, usada para amostrar o valor da resposta. Nesse caso, $f(.)$ precisa ser determinística para que seja possível estimar os parâmetros do modelo. Um exemplo desse tipo de modelo é o modelo logístico:
\begin{equation}\label{eq:logistic}
\Pr(Y_i = 1|\bx_i)= \frac{1}{1+\exp(-\bx_i^\top\bbeta)}.
\end{equation}

\subsection{Modelo \MODEL: um novo modelo baseado em limiar de saturação}

No modelo \emph{fractional q-neighborhood response} de~\cite{Ugander:2013vc}, um indivíduo é considerado ``exposto'' a um tratamento (A ou B) se ele e uma fração maior que $q$ de seus vizinhos for atribuída am mesmo tratamento. A suposição implícita feita por este modelo é de que existe um limiar de saturação: uma vez que um nodo $v$ é considerado ``exposto'', aumentar a fração de vizinhos atribuídos ao mesmo tratamento não afeta a distribuição da resposta $Y_i$. Este modelo não pode ser usado para geração de respostas sintéticas, pois não especifica a distribuição das respostas (seja o nodo exposto ou não).

Nesta seção, propomos o modelo de resposta \MODEL~baseado na suposição da existência de um limiar de saturação $\tau \geq 0.5$. Seja $g_i(\bZ)$ a fração de vizinhos do nodo $i$ no grupo de tratamento. A resposta de um indivíduo é dada por
\begin{equation}\label{eq:tau-exposure}
Y_i = \epsilon_i +
\begin{cases}
\beta_0& \text{se $Z_i = 0$ e $g_i(\bZ) \leq 1-\tau$,}\\
\beta_0 + \beta_2 (g_i(\bZ) - (1-\tau)) & \text{se $Z_i = 0$ e $g_i(\bZ) > 1-\tau$,} \\
\beta_0 + \beta_1 & \text{se $Z_i = 1$ e $g_i(\bZ) \geq \tau$,}\\
\beta_0 + \beta_1 + \beta_2 (g_i(\bZ) - \tau) & \text{se $Z_i = 1$ e $g_i(\bZ) < \tau$.}
\end{cases}
\end{equation}
Para que o modelo seja realista, é necessário que $\beta_1 \beta_2 > 0$, ou seja, que impacto do tratamento do nodo e daquele dos seus vizinhos na resposta esperada tenha o mesmo sinal. Além disso, é necessário que $|\beta_2 \tau| \leq |\beta_1|$ para que as curvas de $E[Y_i]$ quando $Z_i = 0$ e $Z_i = 1$ não se cruzem.

\noindent {\bf Modelo $\tau$-exposure com respostas binárias.} As respostas ($Y_i \in \mathbb{R}$) deste modelo podem ser transformadas em respostas binárias de maneira similar ao modelo probit ou ao modelo logístico. Neste artigo, para a geração de respostas binárias, iremos considerar o sinal de $Y_i$, assim como é feito no modelo probit.

\section{Efeito Médio do Tratamento}

Novamente, assumimos que $Y_i$ é uma variável aleatória condicionada em $\bZ$.
Neste caso, o efeito médio do tratamento ({\it average treatment effect} ou ATE) é o valor esperado da diferença em médias ({\it difference in-means}) entre a resposta de uma unidade quando a população está em tratamento (i.e., $\bZ = \bone$) e a resposta da unidade quando a população está em controle (i.e., $\bZ = \bzero$):
\begin{equation}\label{eq:ate}
\textrm{ATE} = \textrm{E}\left[\frac{1}{N} \sum_{i=1}^N Y_i(\bZ = \bone) - \frac{1}{N} \sum_{i=1}^N Y_i(\bZ = \bzero) \right].
\end{equation}
Esta equação pode ser especializada para cada um dos modelos apresentados na Seção~\ref{sec:response}. Suponha que $g_i(\bZ)$ seja a fração de vizinhos de $i$ em tratamento dado $\bZ$. Neste caso, $g_i(\bZ = \bone) = 1$ and $g_i(\bZ = \bzero) = 0$, for all $i=1,\ldots,N$.

Para o modelo linear~\eqref{eq:linear}, temos
\begin{equation}\label{eq:ate_linear}
\textrm{ATE}_\textrm{linear} = \beta_0 +\beta_1 + \beta_2 - \beta_0 = \beta_1 + \beta_2.
\end{equation}
O teorema de Gauss-Markov diz que os coeficientes $\widehat{\bm{\beta}}$ obtidos pelo método dos quadrados mínimos
são estimadores não-enviesados de mínima variância (MVUE) para $\bm{\beta}$.

Para o modelo probit~\eqref{eq:probit}, temos
\begin{eqnarray}\label{eq:ate_probit}
\textrm{ATE}_\textrm{probit} & = & \Pr(Y_i = 1|\bZ = \bone ) - \Pr(Y_i = 1|\bZ = \bzero ) \nonumber \\
& = & \Phi(\beta_0 + \beta_1 + \beta_2) - \Phi(\beta_0). \label{eq:ate_probit}
\end{eqnarray}

Para o modelo logit ou logístico~\eqref{eq:logistic}, temos
\begin{eqnarray}\label{eq:ate_logit}
\textrm{ATE}_\textrm{logit} & = & \Pr(Y_i = 1|\bZ = \bone ) - \Pr(Y_i = 1|\bZ = \bzero ) \nonumber \\
& = & \frac{e^{-\beta_0} - e^{-(\beta_0 + \beta_1 + \beta_2)} }{(1+e^{-\beta_0})(1+e^{-(\beta_0 + \beta_1 + \beta_2)})}. \label{eq:ate_logistic}
\end{eqnarray}

Para o modelo $\tau$-exposure~\eqref{eq:tau-exposure}, temos
\begin{equation}\label{eq:ate_tau}
\textrm{ATE}_\tau = \beta_0 +\beta_1 - \beta_0 = \beta_1.
\end{equation}

Finalmente, para o modelo $\tau$-exposure binário, temos
\begin{equation}\label{eq:ate_tau}
\textrm{ATE}_{\tau\text{-bin}} = \Phi(\beta_0 +\beta_1) - \Phi(\beta_0).
\end{equation}

\section{Estimadores}\label{sec:estimadores}

\noindent{\bf Modelos linear, probit e logístico.} No caso dos modelos~\eqref{eq:linear},~\eqref{eq:probit} e~\eqref{eq:logistic}, estimadores do ATE podem ser obtidos substituindo-se $\bm{\beta}$ por estimativas de máxima verossimilhança $\widehat{\bm{\beta}}_\text{MLE}$ nas fórmulas \eqref{eq:ate_linear}, \eqref{eq:ate_probit} e \eqref{eq:ate_logit}, respectivamente. Em particular para o modelo linear, o MLE é não-tendencioso e pode ser obtido de maneira mais eficiente através do método dos mínimos quadrados. Já os MLEs dos modelos probit e logístico são apenas consistentes, isto é, convergem para $\bbeta$ quando o número de amostras cresce.
%No caso do modelo linear, Sabe-se que o estimador de mínimos quadrados é não-tendencioso para o modelo~\eqref{eq:linear}, mas é apenas consistente para os modelos~\eqref{eq:probit} e \eqref{eq:logistic}. Um estimador consistente é aquele que converge para o valor real no limite quando o número de amostras tende a infinito (i.e., assintoticamente não-tendencioso).

\noindent{\bf Modelo $\tau$-exposure}. O modelo $\tau$-exposure~\eqref{eq:tau-exposure} pode ser visto como uma regressão linear onde as observações estão divididas em quatro conjuntos:
\begin{eqnarray*}
\mathcal{C}_0 & = & \{i | Z_i = 0 \wedge g_i(\bZ) \leq 1-\tau\}, \\
\overline{\mathcal{C}_0} & = & \{i | Z_i = 0 \wedge g_i(\bZ) > 1-\tau\}, \\
\mathcal{C}_1 & = & \{i | Z_i = 1 \wedge g_i(\bZ) \geq \tau\}, \\
\overline{\mathcal{C}_1} & = & \{i | Z_i = 1 \wedge g_i(\bZ) < \tau\}.
\end{eqnarray*}
A partir desses conjuntos, pode-se construir uma matriz $\bX_{N\times 3}$ onde
\begin{equation}\label{eq:design_tau}
\bx_i = 
\begin{cases}
(1,\, 0,\, 0) & \text{se $i \in \mathcal{C}_0$,}\\
(1,\, 0,\, g_i(\bZ) - (1-\tau) ) & \text{se $i \in \overline{\mathcal{C}_0}$,} \\
(1,\, 1,\, 0) & \text{se $i \in \mathcal{C}_1$,} \\
(1,\, 1,\, g_i(\bZ) - (1-\tau) ) & \text{se $i \in \overline{\mathcal{C}_1}$.} \\
\end{cases}
\end{equation}
Como em qualquer modelo de regressão linear, o estimador de quadrados mínimos dos parâmetros do modelo $\tau$-exposure é dado por $\hat \bbeta_\tau = (\bX^\top \bX)^{-1} \bX \mathbf{y}$. Um estimador não-tendencioso para o ATE pode ser obtido substituindo-se $\bbeta$ por $\hat \bbeta_\tau$ na Equação~\eqref{eq:ate_tau}.
Quando não se deseja supor que a função de resposta é linear na fração de vizinhos em tratamento para os indivíduos em $\overline{\mathcal{C}_0}$  e $\overline{\mathcal{C}_1}$, pode-se usar um estimador mais simples, conhecido como diferença entre médias ({\em difference-in-means}). Ele é definido por
\begin{equation}\label{eq:ate_tau}
\widehat{\text{ATE}}_\tau = \frac{\sum_{i \in \mathcal{C}_1} Y_i}{|\mathcal{C}_1|} - \frac{\sum_{j \in \mathcal{C}_0} Y_j}{|\mathcal{C}_0|}.
\end{equation}
Quando as respostas são oriundas do modelo $\tau$-exposure, o MSE do estimador $\widehat{\text{ATE}}_\tau$ é igual a sua variância, visto que ele é não-enviesado:
\begin{eqnarray}
\MSE(\widehat{\text{ATE}}_\tau) & = & \text{var}
\left(\frac{1}{|\mathcal{C}_1|} \sum_{i \in \mathcal{C}_1} Y_i -
\frac{1}{|\mathcal{C}_0|} \sum_{j \in \mathcal{C}_0} Y_j
\right)  \label{eq:mse_tau} \\
& = & \frac{1}{|\mathcal{C}_1|^2} \text{var}\left(\sum_{i \in \mathcal{C}_1} Y_i \right) +
\frac{1}{|\mathcal{C}_0|^2} \text{var}\left(\sum_{j \in \mathcal{C}_0} Y_j \right)  \nonumber \\
& = & \frac{\sigma^2}{|\mathcal{C}_1|}+\frac{\sigma^2}{|\mathcal{C}_0|}. \nonumber
\end{eqnarray}

\noindent{\bf Modelo $\tau$-exposure binário}. Assim como no modelo $\tau$-exposure, podemos definir o estimador diferença das médias~\eqref{eq:ate_tau}, substituindo $Y_i$ por $Y_i' = \text{sinal}(Y_i)$. No entanto, a expressão para o MSE será diferente, pois $Y_i'$ é uma variável Bernoulli. Quando $i \in \mathcal{C}_1$, temos que $\text{Pr}(Y_i' = 1) = \text{Pr}(Y_i > 0) = \Phi(\beta_0 + \beta_1)$. Quando $i \in \mathcal{C}_0$, temos que $\text{Pr}(Y_i' = 1) = \Phi(\beta_0)$. Logo, a variância de $Y_i'$ é dada por
\begin{equation}\label{eq:var_taubin}
\var(Y_i) = \begin{cases}
\Phi(\beta_0 + \beta_1)(1-\Phi(\beta_0 + \beta_1)) & \text{se $i \in \mathcal{C}_1$,}\\
\Phi(\beta_0)(1-\Phi(\beta_0)) & \text{se $i \in \mathcal{C}_0$.}
\end{cases}
\end{equation}
Substituindo~\eqref{eq:var_taubin} em~\eqref{eq:mse_tau}, temos
\begin{equation}
\MSE(\widehat{\text{ATE}}_{\tau\text{-bin}}) = \frac{\Phi(\beta_0+\beta_1)(1-\Phi(\beta_0+\beta_1))}{|\mathcal{C}_1|} + \frac{\Phi(\beta_0)(1-\Phi(\beta_0))}{|\mathcal{C}_0|}.
\end{equation}

\noindent{\bf SUTVA.} Quando assume-se que a resposta de um indivíduo é influenciada pelo seu tratamento, mas não pelo tratamento atribuído a outros indivíduos, o seguinte estimador {\it difference-in-means} é utilizado para estimar o ATE:
\begin{equation}\label{eq:ate_sutva}
\widehat{\text{ATE}}_\text{SUTVA} = \frac{\sum_{i:Z_i = 1} Y_i}{|\{i:Z_i = 1\}|} - \frac{\sum_{j:Z_j = 0} Y_j}{|\{j:Z_j = 0\}|}.
\end{equation}
Note que~\eqref{eq:ate_sutva} pode ser obtida a partir de~\eqref{eq:ate_tau} definindo-se $\mathcal{C}_1 = \{i:Z_i = 1\}$ e $\mathcal{C}_0 = \{i:Z_i = 0\}$. Nesse caso, o MSE de $\widehat{\text{ATE}}_\text{SUTVA}$ será dado por
\begin{equation*}
\MSE(\widehat{\text{ATE}}_\text{SUTVA}) = \frac{1}{|\mathcal{C}_1|^2} \text{var}\left(\sum_{i \in \mathcal{C}_1} Y_i \right) +
\frac{1}{|\mathcal{C}_0|^2} \text{var}\left(\sum_{j \in \mathcal{C}_0} Y_j \right).
\end{equation*}

\section{Limites inferiores para o erro de estimadores não-enviesados}\label{sec:limites}

Nesta seção derivamos limites inferiores (alguns assintóticos) para o erro de qualquer estimador não-enviesado do ATE, para cada um dos modelos de resposta.
Para isto, usamos o {\it Cram\'er-Rao Lower Bound}  (CRLB), que relaciona o erro médio quadrático (MSE) com a quantidade de informação contida nos dados à respeito dos parâmetros a serem estimados, medida pela matriz de informação de Fisher (FIM). A seguir, denotamos por $T_1(\bX)$, $T_2(\bX)$, $T_3(\bX)$ e $T_4(\bX)$ estimadores não-enviesados de $\text{ATE}_\text{linear}$, $\text{ATE}_\text{probit}$, $\text{ATE}_\text{logit}$ e $\text{ATE}_\tau$, respectivamente. \techreport{As demonstrações podem ser encontradas no Apêndice.}{As demonstrações podem ser encontradas no nosso relatório técnico~\cite{XX}.}

\begin{theorem} \label{th:linear}
O limite inferior do erro de estimação para o modelo linear~\eqref{eq:linear} é dado por
\begin{equation}
\MSE(T_1(\bX)) \geq [0\ 1\ 1] \sigma^2 (\bX^\top \bX)^{-1} [0\ 1\ 1]^\top. \label{eq:crb_linear}
\end{equation}
Note que o limite~\eqref{eq:crb_linear} é válido mesmo que o termo da variância $\sigma^2$ seja desconhecido.
\end{theorem}

No caso do modelo probit, foi provado em~\cite{Demidenko2001} que a informação contida em uma amostra sobre $\bbeta$, medida pela matriz de informação de Fisher (FIM), é dada por 
\begin{equation}
\mathcal{I}(\bX) = \sum_{i=1}^N \frac{\phi^2(s_i)}{\Phi(S_i)(1-\Phi(s_i))} \bx_i \bx_i^\top,
\end{equation}
onde $s_i = \bx_i^\top \bbeta$. Com isso, podemos provar o Teorema~\ref{th:probit}.
\begin{theorem} \label{th:probit}
O limite inferior \emph{assintótico} do erro de estimação para o modelo probit~\eqref{eq:probit} é dado por
\begin{eqnarray}
\MSE(T_2(\bX)) & \geq & (\nabla_{\bm{\beta}}h)^\top \mathcal{I}^{-1}(\bX) \nabla_{\bm{\beta}}h
\end{eqnarray}
onde
$
\nabla_{\bm{\beta}}h = \left(\phi(\bbeta^\top \bone)-\phi(\beta_0), \phi(\bbeta^\top \bone), \phi(\bbeta^\top \bone) \right).
$
\end{theorem}

Segundo~\cite{StatAcumen}, a FIM do modelo logístico é dada por
\begin{equation}\label{eq:logitFIM}
\mathcal{I}(\bX) = \bX^\top \mathbf{W} \bX,
\end{equation}
onde $\mathbf{W} = [W_{ij}]_{N \times N}$ é a matriz diagonal tal que $W_{ii} = \exp(s_i)/(1+\exp(s_i))^2$. No Apêndice, usamos~\eqref{eq:logitFIM} para provar o seguinte teorema.

\begin{theorem} \label{th:logit}
O limite inferior \emph{assintótico} do erro de estimação para o modelo logit~\eqref{eq:logistic} é dado por
\begin{eqnarray}
\MSE(T_3(\bX)) & \geq & (\nabla_{\bm{\beta}}g)^\top \mathcal{I}^{-1}(\bX) \nabla_{\bm{\beta}}g,
\end{eqnarray}
onde
\begin{equation*}
\nabla_{\bm{\beta}}g = \Bigg(\frac{1}{e^{\bbeta^\top \bone}+1} - \frac{1}{(e^{\bbeta^\top \bone}+1)^2 }- \frac{e^{\beta_0}}{(e^{\beta_0}+1)^2},
\frac{e^{\bbeta^\top \bone}}{(e^{\bbeta^\top \bone}+1)^2 }, 
\frac{e^{\bbeta^\top \bone}}{(e^{\bbeta^\top \bone}+1)^2 } \Bigg).
\end{equation*}
\end{theorem}

\begin{theorem} \label{th:tau}
O limite inferior do erro de estimação para o modelo $\tau$-exposure é dado por
\begin{equation}\label{eq:tauMSE}
\MSE(T_4(\bX)) \geq [0\ 1\ 1] \sigma^2 (\bX^\top \bX)^{-1} [0\ 1\ 1]^\top.
\end{equation}
onde a matrix $\bX$ é definida por~\eqref{eq:design_tau}.
\end{theorem}

\section{Estudo empírico sobre a especificação incorreta do modelo}

Nesta seção realizamos um estudo empírico do erro resultante ao se assumir, durante a estimação, um modelo de interferência incorreto. Para tanto, geraremos respostas com os modelos apresentados anteriormente e estudaremos o valor do Erro Quadrático Médio (MSE) ao estimar com os diferentes estimadores, incluindo o MLE do modelo correto.

\noindent{\bf Datasets.} Utilizamos três redes sociais (nodos representam indivíduos) disponíveis na coleção de datasets SNAP, de Stanford \footnote{http://snap.stanford.edu/}.
A Tabela~\ref{tab:redes} mostra as estatísticas básicas delas:
%. São elas \fm{Ana: descrever os datasets}:
\begin{itemize}
\item {\bf Bitcoin OTC:} Rede de pessoas que fizeram transações utilizando o Bitcoin. Uma aresta direcionada $(i, j)$ indica que $i$ confia em $j$ (\textit {who-trusts-whom}).
\item {\bf Enron Email:} Rede formada a partir de cerca 500 mil emails trocados dentro da companhia Enron. Nodos são endereços de email e arestas não-direcionadas $(i,j)$ indicam que pelo menos um email foi enviado de $i$ para $j$ ou de $j$ para $i$. 
\item {\bf Wiki-Vote:} Rede que representa a votação entre colaboradores da Wikipedia para a escolha de administradores das páginas. Nodos na rede representam colaboradores e uma aresta direcionada $(i,j)$ indica que o usuário $i$ votou no usuário $j$.
\end{itemize}

\begin{table}[t]
\footnotesize
\centering
\begin{tabular}{|c|c|c|}
\hline
Dataset& Total de Nodos & Total de Arestas\\
\hline
Bitcoin OTC& 5.881& 35.592\\
\hline
Enron Email&36.692 &183.831\\ 
\hline
Wiki-vote&7.115&103.689\\ 
\hline
\end{tabular}
\caption{Total de nodos e arestas das redes em cada dataset.}
\label{tab:redes}
\end{table}

\noindent{\bf Parâmetros.} Cada um dos modelos é parametrizado por um vetor $\bbeta = (\beta_0, \beta_1, \beta_2)$ cujas coordenadas tem uma interpretação semelhante. O parâmetro $\beta_0$ quantifica a propensão intrínseca da população e será fixado em zero. O parâmetro $\beta_1$ quantifica a influência do tratamento atribuído ao indivíduo na sua resposta. Finalmente, o parâmetro $\beta_2$ quantifica a influência da fração dos vizinhos de um nodo em tratamento na sua resposta. Consideramos os seguintes casos:
\begin{itemize}
\item SUTVA é válida: $\bbeta = (0,1,0)$;
\item tratamento do nodo é irrelevante: $\bbeta = (0,0,1)$;
\item tratamento dos vizinhos é menos relevante que do nodo: $\bbeta = (0,1,0.5)$;
\item tratamento dos vizinhos é tão relevante quanto do nodo: $\bbeta = (0,1,1)$; e
\item tratamento dos vizinhos é mais relevante que do nodo: $\bbeta = (0,1,2)$.
\end{itemize}
Neste trabalho, o parâmetro adicional do modelo $\tau$-exposure será fixado em $\tau = 0.85$.

\noindent{\bf Vetores de tratamento.} Para cada combinação de grafo, vetor $\bbeta$ e modelo de resposta, geramos um vetor de tratamento $\mathbf{Z}$ aleatório, em que cada nodo tinha a mesma probabilidade de pertencer a um dos dois tratamentos.

\noindent{\bf Vetores de resposta.} Para cada vetor de tratamento, geramos 1000 vetores de resposta $\mathbf{Y}$ para um dado modelo.

\noindent{\bf Estimadores.} Para cada par $(\mathbf{Z}, \mathbf{Y})$, calculamos diversas estimativas para o ATE. Usamos o estimador de mínimos quadrados do modelo linear para estimar o ATE nos casos em que a resposta é real (i.e., $Y_i \in \mathbb{R}$). Usamos o MLE dos modelos probit e logístico para estimar o ATE nos casos em que a resposta é binária (i.e., $Y_i \in \{0,1\}$). Além disso, os estimadores $\widehat{\text{ATE}}_\text{SUTVA}$ e $\widehat{\text{ATE}}_\tau$ foram aplicados a todos os pares $(\mathbf{Z}, \mathbf{Y})$. 

\begin{figure}[t]
    \center
    \includegraphics[width=0.8\textwidth]{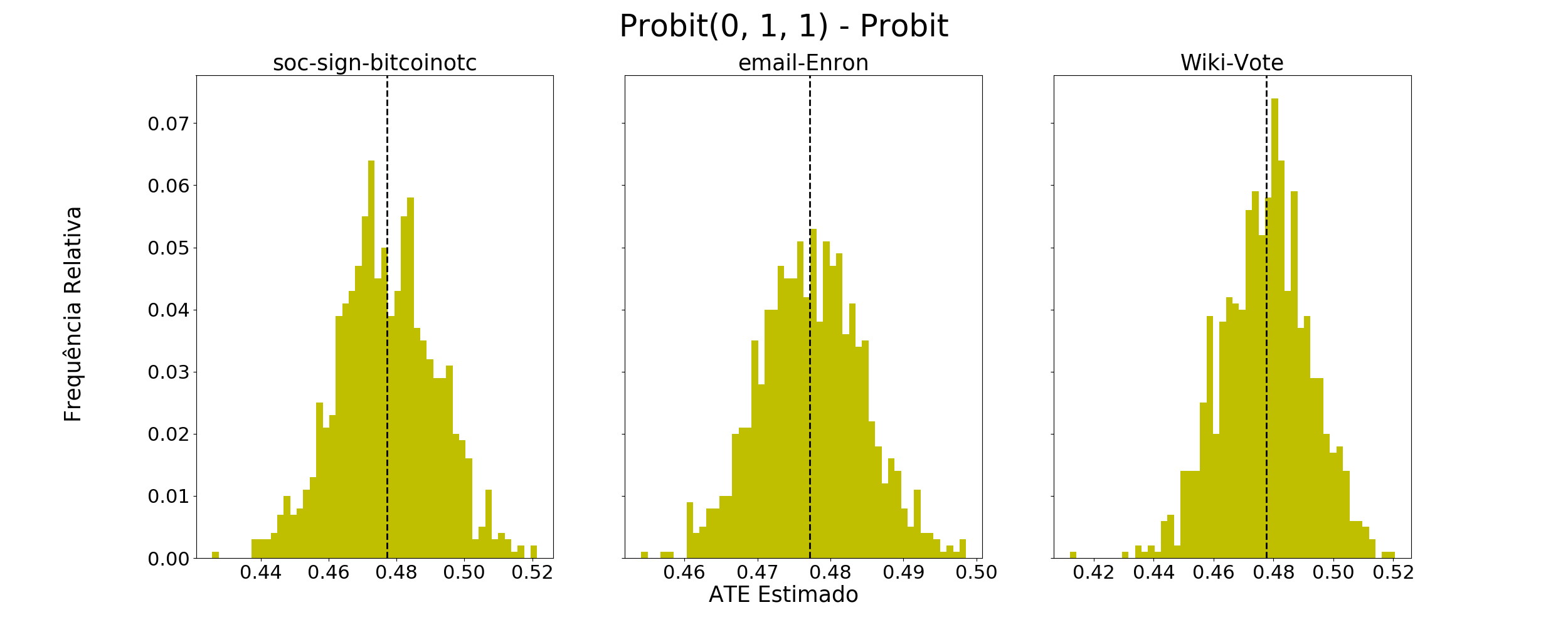}
%     \caption{Distribuição das estimativas $\widehat{\text{\ATE}}_\text{probit}$ obtidas a partir de respostas geradas pelo modelo Probit com $\bbeta = (0, 0, 1)$, para as três redes. a distribuição das estimativas é aproximadamente Normal, com media igual ao ATE real (reta tracejada).}
\caption{Distribuição das estimativas $\widehat{\text{ATE}}_\text{probit}$ obtidas para o modelo Probit com $\bbeta = (0, 1, 1)$, para as três redes. A distribuição é aproximadamente Normal, com media igual ao ATE real (reta tracejada).}
        \label{fig:Hist}
\end{figure}
A Figura~\ref{fig:Hist} mostra as distribuições da estimativa $\widehat{\text{ATE}}_\text{probit}$ \eqref{eq:ate_probit} obtidas a partir de respostas geradas pelo modelo Probit \eqref{eq:probit} com $\bbeta = (0, 1, 1)$, para as três redes reais usadas neste trabalho. Podemos observar que a distribuição das estimativas é aproximadamente Normal. As médias empíricas foram, respectivamente, $0.4772$, $0.4772$, $0.4776$, enquanto o ATE real é cerca de $0.4770$, $0.4774$ e $0.4774$.

Através de inspeção visual, notamos que a distribuição das estimativas para todos os outros experimentos também se assemelha a uma Normal. Embora a média nem sempre seja igual ao ATE real, sempre está muito próxima. Por esta razão, iremos usar o MSE como a estatística que sumariza o resultado de um experimento.

\begin{figure}[ht!]
    \center
    \includegraphics[height=0.76\textheight]{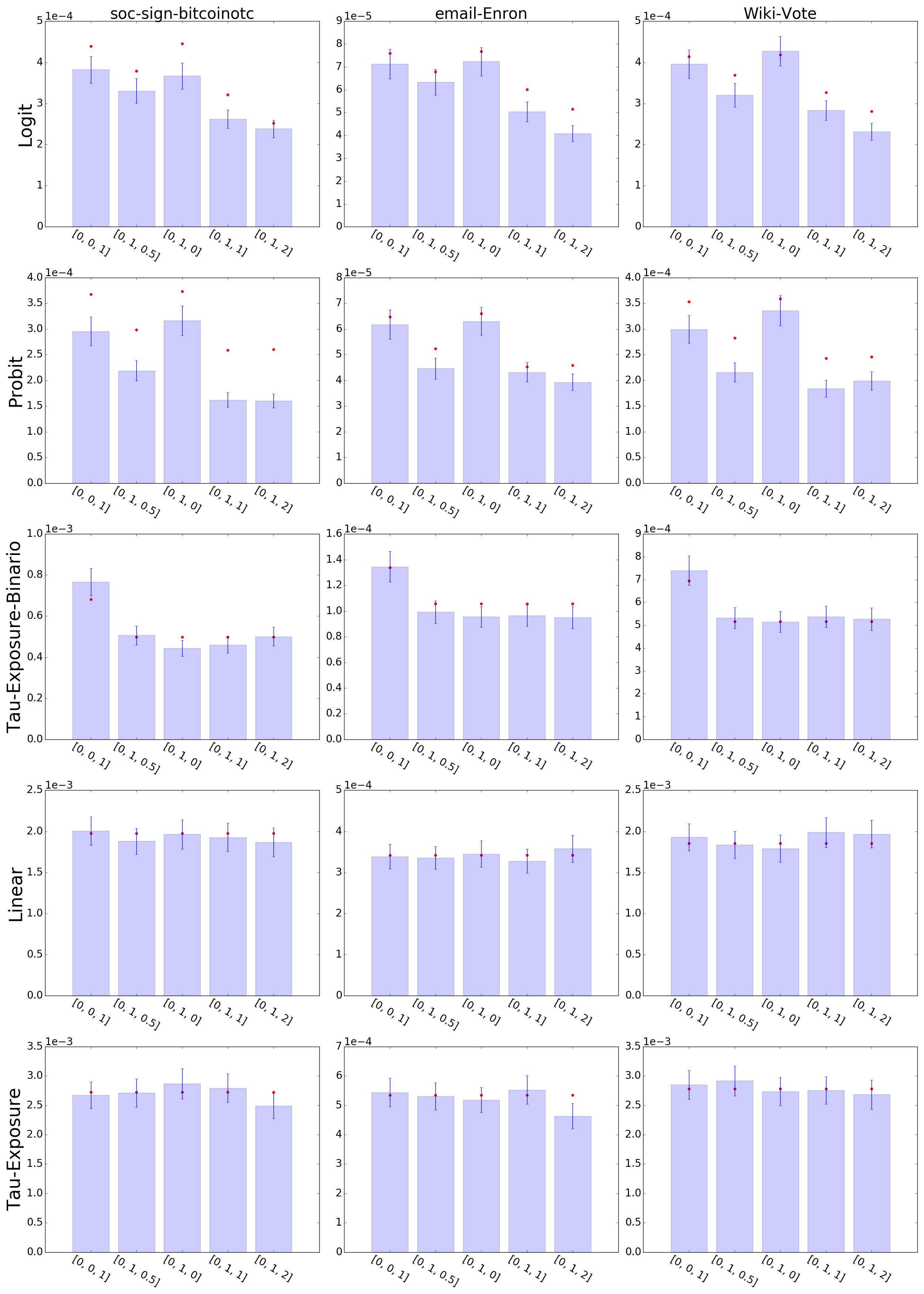}
%     \caption{Distribuição das estimativas $\widehat{\text{\ATE}}_\text{probit}$ obtidas a partir de respostas geradas pelo modelo Probit com $\bbeta = (0, 0, 1)$, para as três redes. a distribuição das estimativas é aproximadamente Normal, com media igual ao ATE real (reta tracejada).}
\caption{Comparação do MSE empírico (barras verticais) com limites inferiores (pontos vermelhos) para estimadores não-enviesados. Com base no intervalo de confiança de 95\%, observa-se que os estimadores enviesados (Logit e Probit) podem resultar em erro menor que o CRLB, e que os não-enviesados atingem o CRLB. }
        \label{fig:Confidence}
\end{figure}

%%%%%% TABELAS MODELOS BINARIOS
A primeira análise que fizemos consiste em comparar os MSEs obtidos ao utilizarmos os estimadores descritos na Seção~\ref{sec:estimadores} com os limites teóricos apresentados na Seção~\ref{sec:limites}. Como os estimadores Probit e Logit são enviesados, é possível que o viés não-nulo seja compensado por uma redução na variância ({\it bias-variance trade-off} \cite{friedman2001elements}). Além disso, os limites inferiores derivados para estes estimadores são assintóticos (i.e., válidos quando o número de amostras tende a infinito). Por outro lado, espera-se que o MSE empírico dos outros estimadores seja maior ou igual aos respectivos CRLBs. A Figura~\ref{fig:Confidence} mostra, para cada dataset e estimador, os resultados do MSE empírico (barras verticais) e os limites inferiores derivados na seção anterior (pontos vermelhos). De fato, com base no intervalo de confiança de 95\%, observa-se que os estimadores enviesados (Logit e Probit) podem resultar em erro menor que o CRLB, e que os não-enviesados atingem o CRLB.

Em seguida, comparamos o MSE obtido ao se escolher o estimador apropriado com aquele obtido ao se escolher o estimador de um modelo diferente. As Tabelas~\ref{tab:logistic} a \ref{tab:tau} mostram os resultados obtidos para cada modelo de resposta usado na geração de respostas sintéticas. Para cada vetor de parâmetros $\bbeta$ em estudo, o menor valor de MSE foi destacado em negrito. Usamos o teste de Welch (equivalente ao teste t para variâncias diferentes) para determinar se a diferença entre o resultado do estimador apropriado e aquele de um estimador incorreto é estatisticamente significativa com $p\text{-value}=0.5$. O asterisco $^*$ indica os casos em que a diferença é significativa.
%As tabelas dos modelos cujas respostas são binárias são apresentados primeiro.

% LOGISTIC
\begin{table}[t]
\footnotesize
\setlength\doublerulesep{0.15cm} 
	\centering
	\begin{tabular}{|l|l|*{5}{c|}}\cline{2-7}
		%\backslashbox{Estimador}{Vetor $\beta$}
		%&\makebox[3em]{$(0,0,1)$}&\makebox[3em]{$(0,1,0.5)$}&\makebox[3em]{$(0,1,0)$}
		%&\makebox[3em]{$(0,1,1)$}&\makebox[3em]{$(0,1,2)$}\\\hline\hline
        \nocell{1} &\multirow{3}{*}{Estimador} & \multicolumn{5}{|c|}{Vetor de parâmetros $\bbeta$ e $\text{ATE}_\text{logistic}$}\\ \cline{3-7}
        \nocell{1} &&$(0,0,1)$ & $(0,1,0.5)$& $(0,1,0)$ & $(0,1,1)$ & $(0,1,2)$\\
        \nocell{1} && $\text{ATE} = 0.23$ & $\text{ATE} = 0.32$ & $\text{ATE} = 0.23$ & $\text{ATE} = 0.38$ & $\text{ATE} = 0.45$ \\ \hline
        \parbox[t]{2mm}{\multirow{4}{*}{\rotatebox[origin=c]{90}{bitcoin}}}
        &SUTVA                &0.05391*               &0.01061*                &\bf{0.00012}*     &0.03416*                &0.08786*\\ \cline{2-7}
		&Logistic             &\bf{0.00037}\phantom{*}&\bf{0.00034}\phantom{*} &0.00040\phantom{*}&0.00027\phantom{*}      &\bf{0.00023}\phantom{*}\\\cline{2-7}
        &Probit               &\bf{0.00037}\phantom{*}&\bf{0.00034}\phantom{*} &0.00040\phantom{*}&\bf{0.00026}\phantom{*} &0.00024\phantom{*}\\\cline{2-7}
		&$\tau$-Exposure      &0.00059*               &0.00048*                &0.00055*          &0.00040*                &0.00032*\\
        \hline \hline
		\parbox[t]{2mm}{\multirow{4}{*}{\rotatebox[origin=c]{90}{Enron}}}
        & SUTVA     &0.05309*                &0.01069*                &\bf{0.00002}*     &0.03503*                &0.09028*\\\cline{2-7}
        & Logistic  &\bf{0.00007}\phantom{*} &\bf{0.00006}\phantom{*} &0.00007\phantom{*}&\bf{0.00005}\phantom{*} &\bf{0.00004}\phantom{*}\\\cline{2-7}
        & Probit    &\bf{0.00007}\phantom{*} &\bf{0.00006}\phantom{*} &0.00007\phantom{*}&\bf{0.00005}\phantom{*} &0.00008*\\\cline{2-7}
        & $\tau$-Exposure &0.00011*          &0.00010*                &0.00011*          &0.00009*                &0.00007*\\
        \hline\hline       
        \parbox[t]{2mm}{\multirow{4}{*}{\rotatebox[origin=c]{90}{Wiki-Vote}}}
        & SUTVA     &0.05421*                &0.01095*               &\bf{0.00010}*     &0.03555*               &0.09181*\\\cline{2-7}
        & Logistic  &\bf{0.00040}\phantom{*} &\bf{0.00032}\phantom{*}&0.00043\phantom{*}&\bf{0.00028}\phantom{*}&\bf{0.00023}\phantom{*}\\\cline{2-7}
        & Probit    &\bf{0.00040}\phantom{*} &\bf{0.00032}\phantom{*}&0.00043\phantom{*}&\bf{0.00028}\phantom{*}&0.00025\phantom{*}\\\cline{2-7}
        & $\tau$-Exposure &0.00063*          &0.00054*               &0.00064*          &0.00047*               &0.00037*\\\hline       
	\end{tabular}
	\caption{MSE do ATE estimado para o modelo de resposta Logistic.}
    \label{tab:logistic}
\end{table}

Conforme esperado, quando a SUTVA é válida (i.e., $\bbeta = (0,1,0)$), o melhor estimador é sempre o SUTVA \eqref{eq:ate_sutva}, independentemente do modelo de resposta. Contudo, o MSE dos demais estimadores não ficou muito maior, nunca mais do que uma ordem de magnitude de diferença. Por outro lado, para outros vetores $\bbeta$, o erro ao se assumir SUTVA foi de até três ordens de magnitude maior. Por exemplo, na Tabela~\ref{tab:logistic}, para o dataset Bitcoin OTC com $\bbeta=(0,1,1)$, o MSE ao se assumir SUTVA foi $0.03416$, ao passo que o MSE do estimador Logistic foi $0.00027$. 

% PROBIT
\begin{table}[t]
\footnotesize
\setlength\doublerulesep{0.15cm} 
	\centering
	\begin{tabular}{|l|l|*{5}{c|}}\cline{2-7}
    
        \nocell{1} &\multirow{3}{*}{Estimador} & \multicolumn{5}{|c|}{Vetor de parâmetros $\bbeta$ e $\text{ATE}_\text{Probit}$}\\ \cline{3-7}
        \nocell{1} &&$(0,0,1)$&$(0,1,0.5)$&$(0,1,0)$&$(0,1,1)$&$(0,1,2)$\\
        \nocell{1} &&$\text{ATE} = 0.34$&$\text{ATE} = 0.43$&$\text{ATE} = 0.34$&$\text{ATE} = 0.48$&$\text{ATE} = 0.50$ \\\hline 
        
        \parbox[t]{2mm}{\multirow{4}{*}{\rotatebox[origin=c]{90}{bitcoin}}}
        &SUTVA    &0.11616 *&0.01699 *&\bf{0.00009} *&0.04729 *&0.09378 *\\\cline{2-7}
		&Logistic &\bf{0.00029} \p&\bf{0.00022} \p&0.00031 \p&0.00017 \p&0.00029 *\\\cline{2-7}
        &Probit   &0.00030 \p&\bf{0.00022} \p&0.00032 \p&\bf{0.00016} \p&\bf{0.00016}\p\\\cline{2-7}
		&\MODEL   &0.00043 *&0.00031 *&0.00046 *&0.00023 *&0.00021 *\\
        \hline \hline
        
		\parbox[t]{2mm}{\multirow{4}{*}{\rotatebox[origin=c]{90}{Enron}}}
        &SUTVA    &0.11711 *&0.01988 *&\bf{0.00001} *&0.05755 *&0.11905 *\\\cline{2-7}
        &Logistic &\bf{0.00006} \p&0.00005 \p&0.00006 \p&0.00009 *&0.00042 *\\\cline{2-7}
        &Probit   &\bf{0.00006} \p&\bf{0.00004} \p&0.00006 \p&\bf{0.00004} \p&\bf{0.00004} \p\\\cline{2-7}
        &\MODEL   &0.00011 *&0.00008 *&0.00010 *&0.00007 *&0.00007 *\\
        \hline\hline 
        
        \parbox[t]{2mm}{\multirow{4}{*}{\rotatebox[origin=c]{90}{Wiki-Vote}}}
        &SUTVA    &0.11254 *&0.01791 *&\bf{0.00008} *&0.05146 *&0.10719 *\\\cline{2-7}
        &Logistic &\bf{0.00030} \p&\bf{0.00022} \p&0.00034 \p&0.00022 *&0.00050 *\\\cline{2-7}
        &Probit   &\bf{0.00030} \p&\bf{0.00022} \p&0.00034 \p&\bf{0.00018} \p&\bf{0.00020} \p\\\cline{2-7}
        &\MODEL   &0.00046 *&0.00036 *&0.00052 *&0.00030 *&0.00028 *\\\hline
        
	\end{tabular}
	\caption{MSE do ATE estimado para o modelo de resposta Probit.}
        \label{tab:probit}
\end{table}

Para o modelo de resposta Logístico \eqref{eq:logistic}, os estimadores Probit e Logístico obtiveram resultados similares, com o segundo apresentando resultados melhores em alguns casos. Por outro lado, observamos na Tabela~\ref{tab:probit} que para o modelo de resposta Probit \eqref{eq:probit}, o estimador Logístico gerou resultados muito ruins para alguns vetores $\bbeta$, em especial para $\bbeta=(0,1,2)$. Observamos também que a diferença do estimador Logístico com o Probit foi menor na menor rede, Bitcoin, e maior na maior rede, Enron.

% TAU-EXPOSURE-BINÁRIO
\begin{table}[t]
\footnotesize
\setlength\doublerulesep{0.15cm} 
	\centering
	\begin{tabular}{|l|l|*{5}{c|}}\cline{2-7}
    
        \nocell{1} &\multirow{3}{*}{Estimador} & \multicolumn{5}{|c|}{Vetor de parâmetros $\bbeta$ e $\text{ATE}_\text{$\tau$-bin}$}\\ \cline{3-7}
        \nocell{1} &&$(0,0,1)$&$(0,1,0.5)$&$(0,1,0)$&$(0,1,1)$&$(0,1,2)$\\
        \nocell{1} &&$\text{ATE} = 0.00$&$\text{ATE} = 0.34$&$\text{ATE} = 0.34$&$\text{ATE} = 0.34$&$\text{ATE} = 0.34$\\\hline 
        
        \parbox[t]{2mm}{\multirow{4}{*}{\rotatebox[origin=c]{90}{bitcoin}}}
        &SUTVA    &0.08506 *&0.01668 *&\bf{0.00009} *&0.07168 *&0.27652 *\\\cline{2-7}
		&Logistic &0.00177 *&\bf{0.00050} \p&0.00029 *&0.00098 *&0.00303 *\\\cline{2-7}
        &Probit   &0.00161 *&0.00054 \p&0.00029 *&0.00107 *&0.00304 *\\\cline{2-7}
		&\MODEL   &\bf{0.00077} \p&0.00051 \p&0.00044 \p&\bf{0.00046} \p&\bf{0.00050} \p\\
        \hline \hline
        
		\parbox[t]{2mm}{\multirow{4}{*}{\rotatebox[origin=c]{90}{Enron}}}
        &SUTVA    &0.08030 *&0.01628 *&\bf{0.00001} *&0.06741 *&0.24642 *\\\cline{2-7}
        &Logistic &0.00193 *&0.00037 *&0.00006 *&0.00133 *&0.00455 *\\\cline{2-7}
        &Probit   &0.00172 *&0.00037 *&0.00006 *&0.00130 *&0.00436 *\\\cline{2-7}
        &\MODEL   &\bf{0.00013} \p&\bf{0.00010} \p&0.00010 \p&\bf{0.00010} \p&\bf{0.00010} \p\\
        \hline\hline 
        
        \parbox[t]{2mm}{\multirow{4}{*}{\rotatebox[origin=c]{90}{Wiki-Vote}}}
        &SUTVA    &0.07951 *&0.01607 *&\bf{0.00008} *&0.06590 *&0.24192 *\\\cline{2-7}
        &Logistic &0.00219 *&0.00060 \p&0.00034 *&0.00156 *&0.00452 *\\\cline{2-7}
        &Probit   &0.00198 *&0.00059 \p&0.00034 *&0.00152 *&0.00432 *\\\cline{2-7}
        &\MODEL   &\bf{0.00074} \p&\bf{0.00053} \p&0.00051 \p&\bf{0.00054} \p&\bf{0.00053} \p\\\hline
        
	\end{tabular}
	\caption{MSE do ATE estimado para o modelo de resposta \MODEL \ Binario.}
            \label{tab:taubin}
\end{table}

Quando o modelo de resposta foi \MODEL\ Binário, nota-se na Tabela~\ref{tab:taubin} que o MSE tanto do estimador logístico quanto do Probit empatam estatisticamente com o do estimador correto quando o efeito do tratamento dos vizinhos é fraco, exceto no dataset Enron, que é a maior rede.

% LINEAR
\begin{table}[t]
\small
\setlength\doublerulesep{0.15cm} 
	\centering
	\begin{tabular}{|l|l|*{5}{c|}}\cline{2-7}
    
        \nocell{1} &\multirow{3}{*}{Estimador} & \multicolumn{5}{|c|}{Vetor de parâmetros $\bbeta$ e $\text{ATE}_\text{Linear}$}\\ \cline{3-7}
        \nocell{1} &&$(0,0,1)$&$(0,1,0.5)$&$(0,1,0)$&$(0,1,1)$&$(0,1,2)$\\
        \nocell{1} &&$\text{ATE} = 1.00$&$\text{ATE} = 1.50$&$\text{ATE} = 1.00$&$\text{ATE} = 2.00$&$\text{ATE} = 3.00$\\\hline 
        
        \parbox[t]{2mm}{\multirow{3}{*}{\rotatebox[origin=c]{90}{bitcoin}}}
        &SUTVA  &0.99433 *&0.24871 *&\bf{0.00069} *&0.99487 *&3.98001 *\\\cline{2-7}
		&Linear &\bf{0.00201} \p&\bf{0.00188} \p&0.00196 \p&\bf{0.00193} \p&\bf{0.00187} \p\\\cline{2-7}
		&\MODEL &0.00275 *&0.00248 *&0.00275 *&0.00274 *&0.00252 *\\
        \hline \hline
        
		\parbox[t]{2mm}{\multirow{3}{*}{\rotatebox[origin=c]{90}{Enron}}}
        &SUTVA  &0.99393 *&0.24829 *&\bf{0.00011} *&0.99373 *&3.97199 *\\\cline{2-7}
        &Linear &\bf{0.00034} \p&\bf{0.00034} \p&0.00035 \p&\bf{0.00033} \p&\bf{0.00036} \p\\\cline{2-7}
        &\MODEL &0.00053 *&0.00053 *&0.00052 *&0.00054 *&0.00061 *\\
        \hline\hline 
        
        \parbox[t]{2mm}{\multirow{3}{*}{\rotatebox[origin=c]{90}{\footnotesize Wiki-Vote}}}
        &SUTVA  &1.00325 *&0.25085 *&\bf{0.00055} *&1.00116 *&4.00782 *\\\cline{2-7}
        &Linear &\bf{0.00193} \p&\bf{0.00184} \p&0.00179 \p&\bf{0.00199} \p&\bf{0.00197}\p\\\cline{2-7}
        &\MODEL &0.00300 *&0.00282 *&0.00280 *&0.00311 *&0.00305 *\\\hline
        
	\end{tabular}
	\caption{MSE do ATE estimado para o modelo de resposta Linear.}
            \label{tab:linear}
\end{table}

Na Tabela~\ref{tab:linear} observamos que, para o modelo de resposta Linear, o erro ao se estimar com \MODEL\  foi em torno de $50$\% maior. Enquanto isso, nota-se na Tabela~\ref{tab:tau} que no modelo \MODEL, o estimador Linear foi até duas ordens de magnitude maior. O SUTVA resultou em MSE até quatro ordens de magnitude maior em ambos os modelos.

% TAU-EXPOSURE
\begin{table}[t]
\small
\setlength\doublerulesep{0.15cm} 
	\centering
	\begin{tabular}{|l|l|*{5}{c|}}\cline{2-7}
    
        \nocell{1} &\multirow{3}{*}{Estimador} & \multicolumn{5}{|c|}{Vetor de parâmetros $\bbeta$ e $\text{ATE}_\text{\MODEL}$}\\ \cline{3-7}
        \nocell{1} &&$(0,0,1)$&$(0,1,0.5)$&$(0,1,0)$&$(0,1,1)$&$(0,1,2)$\\
        \nocell{1} &&$\text{ATE} = 0.00$&$\text{ATE} = 1.00$&$\text{ATE} = 1.00$&$\text{ATE} = 1.00$&$\text{ATE} = 1.00$\\\hline 
        
        \parbox[t]{2mm}{\multirow{3}{*}{\rotatebox[origin=c]{90}{bitcoin}}}
        &SUTVA  &0.57656 *&0.14455 *&\bf{0.00071} *&0.57812 *&2.31417 *\\\cline{2-7}
		&Linear &0.00903 *&0.00368 *&0.00195 *&0.00892 *&0.02943 *\\\cline{2-7}
		&\MODEL &\bf{0.00054} \p&\bf{0.00053} \p&0.00052 \p&\bf{0.00055} \p&\bf{0.00046} \p\\
        \hline \hline
        
		\parbox[t]{2mm}{\multirow{3}{*}{\rotatebox[origin=c]{90}{Enron}}}
        &SUTVA  &0.58681 *&0.14709 *&\bf{0.00010} *&0.58568 *&2.34655 *\\\cline{2-7}
        &Linear &0.01165 *&0.00307 *&0.00034 *&0.01154 *&0.04540 *\\\cline{2-7}
        &\MODEL &\bf{0.00054} \p&\bf{0.00053} \p&0.00052 \p&\bf{0.00055} \p&\bf{0.00046} \p\\
        \hline\hline 
        
        \parbox[t]{2mm}{\multirow{3}{*}{\rotatebox[origin=c]{90}{\footnotesize Wiki-Vote}}}
        &SUTVA  &0.56194 *&0.14214 *&\bf{0.00057} *&0.56368 *&2.24670 *\\\cline{2-7}
        &Linear &0.01283 *&0.00448 *&0.00175 *&0.01225 *&0.04332 *\\\cline{2-7}
        &\MODEL &\bf{0.00285} \p&\bf{0.00292} \p&0.00274 \p&\bf{0.00276} \p&\bf{0.00268} \p\\\hline
        
	\end{tabular}
	\caption{MSE do ATE estimado para o modelo de resposta \MODEL.}
        \label{tab:tau}
\end{table}

\section{Conclusões}

Embora já exista um método para aceitar ou rejeitar a SUTVA \cite{Saveski2017}, não existe um método para determinar qual modelo de resposta melhor descreve os dados. Quando a SUTVA é rejeitada, é preciso assumir um modelo para estimar o ATE. Mesmo que o modelo assumido descrevesse perfeitamente a função de resposta dos usuários, existe um erro inerente a flutuações estatísticas, que derivamos analiticamente neste trabalho. Usando como referência este erro inerente, avaliamos os erros obtidos ao se especificar incorretamente o estimador. Observamos que alguns erros de especificação não elevaram muito o MSE (p.\ ex., assumir Probit, quando o modelo de resposta é Logístico, ou ainda, assumir o $\tau$-exposure com $\tau$ elevado, quando o modelo de resposta é Linear). Contudo, o erro depende, em geral, da rede e dos parâmetros do modelo de resposta.

 \bibliographystyle{sbc}
 \bibliography{networkab}

\techreport{
\titleformat{\section}{\large\bfseries}{\appendixname~\thesection .}{0.5em}{}
\begin{appendix}
\section{Provas dos limites inferiores}

A seguir, derivamos os limites inferiores para o erro de estimação do ATE para cada modelo de resposta.
\begin{proof}[Prova do Teorema~\ref{th:linear}]
Considere o modelo linear de resposta descrito por~\eqref{eq:linear}. Segundo o teorema de Gauss-Markov, os coeficientes $\hat \bbeta$ são estimadores não-enviesados de mínima variância (MVUE) de $\bbeta$. A matriz de covariância de $\hat \bbeta$ é dada por $\sigma^2 (\bX^\top \bX)^{-1}$. Logo, o MSE do estimador $\widehat{\text{ATE}}_\text{linear} = \hat \beta_1 + \hat \beta_2$ é dado por
\begin{eqnarray*}
	\text{MSE}(\widehat{\text{ATE}}_\text{linear}) & = & \text{var}(\hat \beta_1 + \hat \beta_2) \\
    & = & \text{var}(\hat \beta_1) + \text{var}(\hat \beta_2) + 2\text{cov}(\hat \beta_1, \hat \beta_2) \\
    & = & [0\ 1\ 1] \sigma^2 (\bX^\top \bX)^{-1} [0\ 1\ 1]^\top. 
\end{eqnarray*}
\end{proof}

Diferente do modelo linear, o ATE dos modelos probit~\eqref{eq:ate_probit} e logit~\eqref{eq:ate_logistic} não é uma função linear dos parâmetros $\bbeta$. Nestes casos, é preciso usar o método Delta~\cite[Capítulo 20]{efron2016computer}, que fornece uma aproximação de primeira ordem para a variância de um estimador
$h(\bT(\bX))$ a partir da matriz de covariância de um estimador $\hat \bbeta = \bT(\bX)$, onde $h$ é uma função não-linear. A aproximação é dada por
\begin{equation}\label{eq:delta}
\text{var}(h(\bT(\bX))) \approx (\nabla_{{\bbeta}}h(\bT(\bX)))^\top  \,\text{cov}(\bT(\bX))\, \nabla_{{\bbeta}} h(\bT(\bX)),
\end{equation}
onde $\nabla_{{\bbeta}} h(\bT(\bX))$ denota o gradiente de $h(\bT(\bX))$ em relação à $\bbeta$ e $\text{cov}(\bT(\bX))$ é a matriz de covariância do estimador $\bT(\bX)$. Pelo teorema Crámer-Rao~\cite[Capítulo 5]{efron2016computer}, a diferença entre a a matriz de covariância de qualquer estimador não-tendencioso $\bT(\bX)$ e a inversa da matriz de informação de Fisher $\mathcal{I}(\bX)$ é positiva semidefinida (i.e., $\text{cov}(\bT(\bX)) - \mathcal{I}^{-1}(\bX) \succeq 0$). Portanto, substituindo $\text{cov}(\bT(\bX))$ por $\mathcal{I}^{-1}(\bX)$ em~\eqref{eq:delta}, obtemos a seguinte aproximação assintótica (quando o número de amostras cresce):
\begin{equation}\label{eq:crlb}
\text{var}(h(\bT(\bX))) \geq (\nabla_{{\bbeta}}h(\bT(\bX)))^\top  \,\mathcal{I}^{-1}(\bX)\, \nabla_{{\bbeta}} h(\bT(\bX)).
\end{equation}

\begin{proof}[Prova do Teorema~\ref{th:probit}]No caso do modelo probit, $\text{ATE}_\text{probit} = h(\bbeta) = \mathbf{\Phi}( \bbeta^\top \bone) - \mathbf{\Phi}( \beta_0)$. Logo,
\begin{eqnarray}
\MSE(\widehat{\text{ATE}}_{\text{probit}}) & \geq & (\nabla_{\bm{\beta}}h)^\top \mathcal{I}^{-1}(\bX) \nabla_{\bm{\hat\beta}}h,
\end{eqnarray}
onde
\begin{eqnarray}
\nabla_{\bm{\beta}}h & = & \nabla_{\bm{\hat\beta}} [\Phi(\bbeta^\top \bone) - \Phi(\beta_0)] \nonumber \\
& = & \left(\phi(\bbeta^\top \bone)-\phi(\beta_0), \phi(\bbeta^\top \bone), \phi(\bbeta^\top \bone) \right).
\end{eqnarray}
\end{proof}

\begin{proof}[Prova do Teorema~\ref{th:logit}]No caso do modelo logístico,  
\begin{equation*}
\text{ATE}_\text{logit} = h( \bbeta) = \frac{e^{-\beta_0} - e^{-(\beta_0 + \beta_1 + \beta_2)} }{(1+e^{-\beta_0})(1+e^{-(\beta_0 + \beta_1 + \beta_2)})}.
\end{equation*}
 Logo,
\begin{eqnarray}
\MSE(\widehat{\text{ATE}}_{\text{logit}}) & \geq & (\nabla_{\bm{\beta}}h)^\top \mathcal{I}^{-1}(\bX) \nabla_{\bm{\beta}}h,
\end{eqnarray}
onde
\begin{eqnarray}
\nabla_{\bm{\beta}}h & = & \nabla_{\bm{\beta}} \left[\frac{e^{-\beta_0} - e^{-(\beta_0 + \beta_1 + \beta_2)} }{(1+e^{-\beta_0})(1+e^{-(\beta_0 + \beta_1 + \beta_2)})} \right] \nonumber \\
& = & \Bigg(\frac{1}{e^{\bbeta^\top \bone}+1} - \frac{1}{(e^{\bbeta^\top \bone}+1)^2 }- \frac{e^{\beta_0}}{(e^{\beta_0}+1)^2}, \frac{e^{\bbeta^\top \bone}}{(e^{\bbeta^\top \bone}+1)^2 }, \frac{e^{\bbeta^\top \bone}}{(e^{\bbeta^\top \bone}+1)^2 } \Bigg).
\end{eqnarray}
\end{proof}

\begin{proof}[Prova do Teorema~\ref{th:tau}] A prova segue imediatamente da observação de que o modelo $\tau$-exposure~\eqref{eq:tau-exposure} pode ser visto como uma regressão linear.
\end{proof}

\end{appendix}
}{}

\end{document}